\documentclass[twocolumn]{aastex631}

\usepackage{amsmath}

\newcommand{\Hb}{\ensuremath{\rm H\beta}}
\newcommand{\Oiii}{[\ion{O}{3}]}
\newcommand{\Oiiiuv}{\ion{O}{3}]}

\newcommand{\Ciiiuv}{\ion{C}{3}]}

\newcommand{\Siii}{[\ion{S}{3}]}

\newcommand{\Hi}{\ion{H}{1}}
\newcommand{\Hii}{\ion{H}{2}}

\newcommand{\Te}{\ensuremath{\rm T_e}}

\begin{document}

\title{Consistent gas-phase C/O abundances from UV and optical emission lines: \\
a robust scale for chemical evolution across cosmic time}

\author[0009-0009-4213-3630]{Paige M. Kelly}
\affiliation{Department of Physics and Astronomy, University of California, Davis, 1 Shields Ave, Davis, CA 95616, USA}

\author[0000-0001-5860-3419]{Tucker Jones}
\affiliation{Department of Physics and Astronomy, University of California, Davis, 1 Shields Ave, Davis, CA 95616, USA}

\author[0000-0003-4520-5395]{Yuguang Chen}
\affiliation{Department of Physics, The Chinese University of Hong Kong, Shatin, N.T., Hong Kong SAR, China}
\affiliation{Department of Physics and Astronomy, University of California, Davis, 1 Shields Ave, Davis, CA 95616, USA}

\author[0000-0003-4792-9119]{Ryan L. Sanders}
\affiliation{Department of Physics and Astronomy, University of Kentucky, 505 Rose Street, Lexington, KY 40506, USA}

\author[0000-0002-4153-053X]{Danielle A. Berg}
\affiliation{Department of Astronomy, The University of Texas at Austin, 2515 Speedway, Stop C1400, Austin, TX 78712, USA}

\author[0000-0002-9132-6561]{Peter Senchyna}
\affiliation{Observatories of the Carnegie Institution for Science, 813 Santa Barbara Street, Pasadena, CA 91101, USA}

\author{Fabio Bresolin}
\affiliation{Institute for Astronomy, University of Hawaii, 2680 Woodlawn Drive, Honolulu, HI 96822, USA}

\author{Daniel Stark}
\affiliation{Department of Astronomy, University of California, 501 Campbell Hall \#3411, Berkeley, CA 94720, USA}




\begin{abstract}

The carbon to oxygen (C/O) abundance ratio is a valuable tracer of star formation history, as C and O enrichment occurs on different timescales. However, measurements based on ultraviolet (UV) collisionally excited lines and those based on optical recombination lines may be subject to biases from the abundance discrepancy factor (ADF), which is well established for oxygen but uncertain for carbon. We present precise UV-based measurements of gas-phase C$^{2+}$/O$^{2+}$ ionic abundance in four \Hii\ regions which have prior optical-based measurements, combined with archival UV data for two additional \Hii\ regions, in order to establish a reliable abundance scale and to investigate biases between the two methods. We find a clear ADF for the C$^{2+}$ ion which is consistent with that of O$^{2+}$, assuming a similar temperature structure in the zones of the nebula which these ions occupy. The C/O abundance derived from UV collisional lines and optical recombination lines is therefore also consistent to within $<0.1$ dex, with an offset of $0.05\pm0.03$ dex in C$^{2+}$/O$^{2+}$ for the standard \Te\ method. While the absolute C/H and O/H abundances are subject to large uncertainty from the ADF, our results establish that C/O abundances measured from these different methods can be reliably compared. Thus we confirm the robustness of gas-phase C/O measurements for studying galaxy evolution and star formation timescales, including from rest-UV observations of high redshift galaxies with JWST.

\end{abstract}



\section{Introduction} \label{sec:intro}

Elements heavier than helium (``metals'') are predominantly formed by stars and thus can give insight into the history of star formation, gas accretion, and outflows which ultimately regulate galaxy formation \citep[e.g.,][]{Maiolino2019,Lilly2013,Finlator2008,Erb2008}. 

The abundance of heavy elements relative to hydrogen (``metallicity'') is a widely used probe, especially the gas-phase oxygen abundance (O/H) which is commonly measured at high redshifts from nebular emission lines \citep[reaching $z\sim10$; e.g.,][]{Sanders2024,Langeroodi2023,Sarkar2025,Kewley2019}.
In addition, the carbon-to-oxygen (C/O) gas-phase abundance ratio can be especially informative since these elements originate from stars of different masses. Oxygen primarily originates from high-mass stars (M $\gtrsim 8 \, {M}_\odot$) which explode as core-collapse supernovae (SNe) within approximately 10 Myr, enriching the gas. Carbon is also released in core-collapse SNe, but it is largely produced in intermediate-mass stars (M~$\sim$~2--5~${M}_\odot$) and released after $\gtrsim$100 Myr during the asymptotic giant branch stage. Because carbon and oxygen originate in stars of different masses and hence different lifetimes, the C/O ratio exhibits variation on short timescales \citep[e.g.][]{Garnett1995, Esteban2014, Jones23, Berg2019}.

In a very young galaxy that has formed the majority of its stellar mass in the past 100 Myr, the C/O abundance ratio will be near the value produced from pure enrichment by core-collapse SNe alone, establishing a “floor” of approximately log(C/O)~$\sim-1.0$ \citep[e.g.,][]{Nomoto2013,Kobayashi2023}. Roughly 100~Myr after a starburst event, AGB stars will start releasing significant amounts of carbon, but relatively little oxygen \citep[e.g.,][]{Kobayashi20}, increasing the C/O ratio. In subsequent starburst events, core-collapse SNe will again release large amounts of oxygen, decreasing the C/O ratio, followed by a later increase in the C/O ratio as new generations of AGB stars release carbon. This time sensitivity of the C/O ratio relative to star formation events, combined with outflows that preferentially remove core-collapse SNe products such as O, results in significant variations in the C/O abundance ratio which can provide powerful inferences on star formation histories \citep[e.g.,][]{Berg2019,Jones23}.

Gas-phase metallicity is often measured from nebular emission lines. However, such studies are subject to a systematic uncertainty known as the abundance discrepancy factor (ADF), which is well established for oxygen abundances. Specifically, O/H measured from recombination lines (RLs) is 0.2--0.3 dex higher than the same quantity measured from collisionally excited lines (CELs) \citep[e.g.,][]{Rojas2007, Esteban2009, Toribio2017}. This discrepancy is commonly attributed to temperature fluctuations \citep[e.g.,][]{Peimbert1967, Peimbert2017}, although an alternative explanation of the ADF is that the RLs may be biased by metallicity inhomogeneities \citep{Torres-Peimbert1990, Stasinska2007, Chen2023}. In a scenario with cool, metal rich, and hydrogen poor clumps the abundances from RLs may be overestimated such that the true abundance would be closer to those derived from CELs. 
However, such a scenario would also include temperature fluctuations which can simultaneously affect the CELs. Abundances derived from RLs are generally considered more reliable because they are not sensitive to temperature. A practical difference is that RLs of heavy elements are far less luminous than CELs, such that metallicity measurements beyond the relatively nearby universe are based entirely on CELs.  

It is routinely assumed that carbon is subject to the same ADF as oxygen \citep[e.g.,][]{Toribio2017,Berg2016,Esteban2014,Jones23}. However, the sample of ADF(C) measurements is limited and generally affected by substantial uncertainties \citep[e.g.,][]{Toribio2017, Esteban2009, Torres1980, Peimbert93, Garnett1995, Peimbert2003, Esteban2004, DelgadoOrion2021}. Nebular carbon and C/O abundances are most readily measured from recombination lines at optical wavelengths or CELs in the ultraviolet (UV). C/O measurements in nearby galaxies and \Hii\ regions are predominantly from optical RLs at high metallicity, and UV CELs at low metallicity (12+log(O/H)~$\lesssim 8$). High redshift C/O measurements are exclusively based on CELs, which are accessible at observed optical wavelengths at $z\gtrsim2$ and have now been probed up to $z>8$ with spectra from the James Webb Space Telescope \citep[JWST; e.g.,][]{Cordova2022,Jones23,Hu24}.
This growing body of C/O measurements based on UV CELs across a wide range of redshifts holds great promise for chemical evolution studies. 
A key goal of this paper is to determine whether these UV CEL-based results are robust and reliable compared to the RL-based C/O abundance measurements of local galaxies -- or whether they are subject to a systematic difference due to the ADF. Ultimately, this will enable accurate chemical evolution studies with C/O, spanning measurements from the reionization epoch to the present-day universe.

Establishing the ADF of carbon and the consistency of UV- and optical-based C/O abundances requires a sample of targets with both CEL and RL measurements. We note that in practice this analysis is based on the C$^{2+}$ and O$^{2+}$ ions, as opposed to the total abundances, although ionization correction factors are often small \citep[e.g.,][]{Berg2016}.
Previously, \citet{Toribio2017} reported the ADF(C$^{2+}$) for 6 \Hii\ regions and found that they are consistent on average with ADF(O$^{2+}$). 
Their CEL-based C/O and C/H measurements required using the ratio of UV \Ciiiuv~$\lambda\lambda$1907,1909 to optical emission lines (e.g., \Oiii~$\lambda$5007 and \Hb). 
Such measurements are highly sensitive to dust attenuation between UV and optical lines, as well as potentially non trivial relative flux calibration between the UV and optical spectra. 
The combination of UV and optical data thus limits the precision of CEL-based C$^{2+}$/O$^{2+}$ measurements.

A goal of our work in this paper is to minimize uncertainty by measuring the C$^{2+}$/O$^{2+}$ ratio using only UV CELs, specifically \Ciiiuv~$\lambda\lambda$1907,1909 and \Oiiiuv~$\lambda\lambda$1661,1666. This reduces the corrections and associated uncertainty for dust attenuation and cross-instrument flux calibration \citep[e.g.,][]{Izotov2023}. These UV CEL-based results allow a direct comparison with the C$^{2+}$/O$^{2+}$ ratio measured using optical RLs. With a previously determined ADF(O$^{2+}$), we are able to find the ADF(C$^{2+}$). There are only two existing reliable measurements of C$^{2+}$ and O$^{2+}$ derived from these two methods with equivalent UV data, and we include them in our analysis. Additionally, we investigate whether the same temperature fluctuations proposed to cause the oxygen ADF can explain that of carbon. The dimensionless variance, t$^{2}$, quantifies the magnitude of the temperature fluctuations. It is typically calculated to bring oxygen abundances derived from CELs into exact agreement with the values determined from RLs. We use this value to correct both carbon and oxygen in the C$^{2+}$/O$^{2+}$ ratio. 

The magnitude of the fluctuations can
be quantified by t2 , the dimensionless variance that is derived from the ADF.

The paper is structured as follows.
We outline the selected \Hii\ regions, observations, and data reduction in Section~\ref{sec:observations}. In Section~\ref{sec:Analysis} we describe our emission line fits, dust attenuation, and sources of uncertainty in the C$^{2+}$/O$^{2+}$ abundance. Finally, we present our C$^{2+}$/O$^{2+}$ measurements determined using UV CELs and temperature fluctuations along with the resulting ADF(C$^{2+}$) in Section~\ref{sec:results}, with the main conclusions discussed in Section~\ref{sec:conclusions}.

\section{Sample Selection, Observations, and Data Reduction}
\label{sec:observations} 

Our direct goal is to measure C$^{2+}$/O$^{2+}$ from UV CELs and compare with C$^{2+}$/O$^{2+}$ from RLs. This allows us to examine the carbon ADF relative to the well established oxygen ADF, and determine the reliability of C/O measurements from the two methods. In this section we describe the target sample and UV spectroscopy which provides these measurements.

\subsection{Objects}
\label{sec:objects}

\begin{table*}

 \centering
 \begin{tabular}{lcccccc}
  \hline
   &NGC $5408$ & $N81$ & $N66A$ & $N44C$ & Mrk $71$ & $N88A$\\
  \hline
  C(H$\beta$)& $0.25\pm0.05$ & $0.11\pm0.07$ & $0.21\pm0.07$ & $0.21\pm0.09$ & $0.12\pm0.04$ &  \\ 
  \Te(\Oiii) [K]& $16000\pm400$ & $12900\pm150$ & $12600\pm100$ & $11400\pm100$ & $16700\pm300$ & $15000\pm200$\\
  \Te(\Siii) [K]& $15700\pm950$ & $13900\pm550$ & $14200\pm550$ & $11400\pm500$ &  & $13900\pm350$\\
  $n_e$ [cm$^{-3}$]& $430\pm300$ & $400\pm100$ & $320\pm100$ & $440\pm150$ & $20^{+430}_{-20}$ & $4300\pm400$\\
  $\log C^{2+}/O^{2+}_{RLs}$ & $-0.84\pm0.15$ & $-0.66\pm0.04$ & $-0.65\pm0.06$ & $-0.43\pm0.03$ & $-0.47 \pm 0.10$ & $-0.55 \pm 0.03$\\
  
  $12+\log(O/H)$ & $8.21 \pm 0.04$ & $8.34\pm0.02$ & $8.35 \pm 0.03$ & $8.58\pm0.02$ & $8.04\pm0.05$ & $8.22\pm0.02$ \\
  ADF(O$^{2+}$) & $0.46\pm0.05$ & $0.33\pm0.11$ & $0.34\pm0.13$ & $0.32\pm0.03$ & $0.34\pm0.05$ & $0.32\pm0.10$\\
  $t^2$ & $0.147 \pm 0.017$ & $ 0.091 \pm 0.017 $ & $ 0.090 \pm 0.019 $ & $0.069 \pm 0.016 $ & $0.120\pm0.022$ & $0.107\pm0.027$\\

  \hline
  $\log C^{2+}/O^{2+}_{CELs}$ & -0.78$\pm0.05$ & -0.73$\pm0.05$ & -0.73$^{+0.07}_{-0.10}$ & -0.58$\pm0.06$ & -0.52$\pm0.04$ & -0.61$^{+0.12}_{-0.18}$ \\
  ADF(C$^{2+}$) & $0.38\pm0.16$ & $0.40\pm0.12$ & $0.41\pm0.16$ & $0.47\pm0.07$ & $ 0.39\pm0.12 $ & $ 0.36^{+0.16}_{-0.23}$ \\
  OIII] $\lambda\lambda$ 1660, 1666  & $12.0\pm0.8$ & $43.4\pm1.3$ & $7.9\pm1.0$ & $7.4\pm0.7$ & $ 11.4\pm 1.0$ & $ 27.3\pm 6.6$ \\
  CIII] $\lambda\lambda$ 1907, 1909  & $42.7\pm1.9 $ & $ 165.6\pm 2.4 $ & $36.6 \pm2.2 $ & $ 39.6\pm2.1 $ & $53.6 \pm 2.1$ & $ 108.0\pm6.0 $\\
  \hline

 \end{tabular}
  \caption{Physical properties of the nebulae in our sample. 
  We list archival measurements of the reddening coefficient C(H$\beta$), electron temperatures (\Te), and density ($n_e$) measured from the [\ion{Cl}{3}] doublet (except for NGC 5408 which is an average). We include the carbon-to-oxygen and oxygen-to-hydrogen gas phase abundances derived from optical RLs, expressed as $C^{2+}$/$O^{2+}_{RLs}$ and $12+\log(O/H)$. The oxygen ADF(O$^{2+}$) and the temperature fluctuation parameter, $t^2$, are also sourced from the literature based on optical spectroscopy. Values for N66A, N44C, N81, and N88A are from \citet{Toribio2017}, those for NGC 5408 are from \citet{Esteban2014}, and values for Mrk 71 are from \citet{Esteban2009}. These values are used in our analysis to calculate $C^{2+}$/$O^{2+}_{CELs}$ and ADF(C$^{2+}$) from UV spectroscopy described in this paper. The fluxes of \Oiiiuv\ and \Ciiiuv\ presented in the lower two rows are in units of 10$^{-15}$~erg$^{-1}$s$^{-1}$cm$^{-2}$. The fluxes for Mrk 71 and N88A are from \citet{Smith23} and \citet{Kurt99}, respectively, while all others are from the analysis in this paper. We recalculated \Te and ADF($O^{2+}$) using the atomic data sets adopted in this work.}
  \label{tab:rls}
\end{table*}

Our targets are drawn from the Hubble Space Telescope (HST) program GO-16697 (PI: R. Sanders), which obtained UV spectroscopy of nine nearby \Hii\ regions with the Cosmic Origins Spectrograph (COS). All the {\it HST} data used in this paper can be found in MAST: \dataset[10.17909/8z20-fy09]{http://dx.doi.org/10.17909/8z20-fy09}. All targets were selected to have previous measurements of C$^{2+}$/O$^{2+}$ from optical RLs \citep{Toribio2017,Esteban2014}. 
We detect the \Oiiiuv~$\lambda\lambda$1661,1666 and \Ciiiuv~$\lambda\lambda$1907,1909 CELs in four objects from which we can reliably determine C$^{2+}$/O$^{2+}$: NGC 5408, N66A, N44C, and N81. Additionally, we include two objects from the literature, Mrk 71 and N88A, which were observed with HST's Faint Object Spectrograph (FOS) in the UV as described in \citet{Garnett1995} and we pull optical results from \citet{Esteban2009} and \citet{Toribio2017}. Relevant properties measured from optical spectroscopy for our sample are listed in Table ~\ref{tab:rls}.

\subsection{HST/COS}
\label{sec:HST} 

HST/COS offers high sensitivity UV spectroscopy, with suitable wavelength coverage and resolution for our objectives. 
COS has two observing channels: a far ultraviolet (FUV) and a near ultraviolet (NUV) detector, which we describe in detail in Sections~\ref{sec:fuv} and ~\ref{sec:nuv}, respectively. We use both channels to observe \Oiiiuv\ (FUV) and \Ciiiuv\ (NUV) emission lines. 
It is important to note that the COS aperture size is 2\farcs5, while the extraction area for the optical spectra used in this analysis is 3$\times$5 arcsec$^{2}$ for NGC 5408, 3$\times$5.3 for N88A, 5.76$\times$1.7 for Mrk 71, and 3$\times$9.4 arcsec$^{2}$ for N81, N66A, and N44C. To minimize aperture effects, the HST/COS aperture covered the brightest knots of emission within the optical spectroscopic aperture of each target, where we expect the ionizing sources and the bulk of the nebular emission to be. 

\subsubsection{COS FUV} \label{sec:fuv} 

The COS FUV detector is a windowless cross delay line (XDL) instrument with two consecutive segments, FUVA and FUVB.  We used the G160M grating with a central wavelength of 1600 {\AA}, which covers wavelengths from 1360 to 1775 {\AA}. The G160M grating was chosen over the  G140L and G230L gratings for its sensitivity and higher spectral resolution, necessary for separating \Oiiiuv~$\lambda$1666 in our targets from \ion{Al}{2}~$\lambda$1671 absorption in the Milky Way.

Our observations were taken at lifetime position (LP) 4. The default extraction technique used after LP3 was switched from BOXCAR to TWOZONE. We tested the difference in flux between a BOXCAR and TWOZONE extraction for our FUV data and found no significant change. These results are similar to those found in \citet{James2022}. Therefore, we extracted our 1D spectra using TWOZONE. 

\subsubsection{COS NUV}
\label{sec:nuv} 
COS has a Multi-Anode Micro-channel Array (MAMA) detector that is sensitive to the wavelengths in the NUV. We used the medium-dispersion grating G185M with three parallel stripes covering 35 {\AA} each with 64 {\AA} between them. 
For our purposes, CENWAVE 1913 was chosen such that \Ciiiuv~$\lambda\lambda$ 1907,1909 is covered in the central stripe. 

The default extraction routine in the COS pipeline is optimized for point sources, while our \Hii\ region targets are extended ($>$ 0\farcs6 FWHM). The extraction technique for data taken with the COS NUV detector required optimization for our extended sources. For all of our objects, the BOXCAR extraction region was increased to include more of the flux, and for many of them, the light distribution from the central stripe overlaps with the other two. 

We visually inspected the 1-D spatial profiles to assess the necessary extraction width in the central stripe and the degree of overlapping flux from other stripes. We find that a height of 95 pixels (as recommended by \citealt{James2022}) provides good results, and adopt this for the spectral extraction of our targets.

\subsection{Flux calibration}
\label{sec:coadding} 

The emission lines of interest are \Ciiiuv~$\lambda\lambda$ 1907, 1909 and \Oiiiuv~$\lambda\lambda$ 1660, 1666, observed using the HST/COS G185M and G160M gratings, respectively. 
To ensure accurate flux ratios, we obtained a relative calibration by fitting a power law through the featureless continuum regions of each spectrum.
We extrapolate the G160M best-fit power law to the wavelength of \Ciiiuv, and scale the local best-fit continuum level in G185M to the same value. The applied scaling factor is $<$15\% in all cases.
The final G160M and G185M spectra are shown in Figure~\ref{fig:fits}.

\begin{figure}
	\includegraphics[width=\columnwidth]{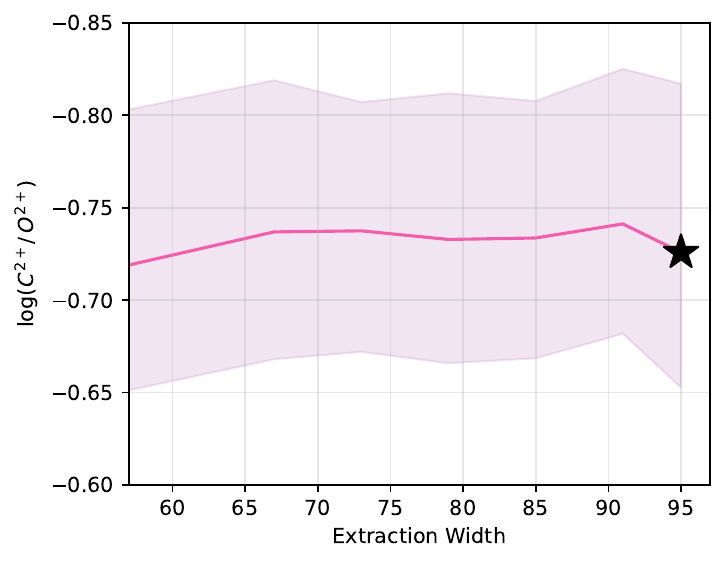}
    \caption { C$^{2+}$/O$^{2+}$ abundance derived for N66A at varying extraction widths of the G185M spectrum, which covers the \Ciiiuv\ emission. 
    The star indicates the chosen BOXCAR pixel height of 95 pixels. This was chosen based on the spatial profile of emission seen in the 2-D spectra, as described in the text. Uncertainty due to the extraction width was determined using C$^{2+}$/O$^{2+}$ at heights from 57 to 95. This corresponds to an uncertainty of $\pm0.01$ dex, which is small compared to the statistical uncertainty (pink shaded region). In all targets we find that uncertainty due to the extraction aperture is small compared to other sources of uncertainty.}
    \label{fig:heights}
\end{figure}

We note that some of the objects are more extended than the adopted extraction widths. 
To assess the effects of spatial extent and extraction width, we tested how the derived C$^{2+}$/O$^{2+}$ changes with varying widths of the BOXCAR extraction region for the G185M grating.
Figure~\ref{fig:heights} shows log(C$^{2+}$/O$^{2+}$) from our analysis as a function the G185M BOXCAR extraction width for an example target N66A. The extraction width does not significantly affect C$^{2+}$/O$^{2+}$ -- we found a maximum difference of 0.03 dex in log(C$^{2+}$/O$^{2+}$) when considering widths from 57-95 pixels.

\section{Analysis}

\label{sec:Analysis} 
\subsection{Emission line fluxes}
\label{sec:fits} 
\begin{figure}
	\includegraphics[width=\columnwidth]{"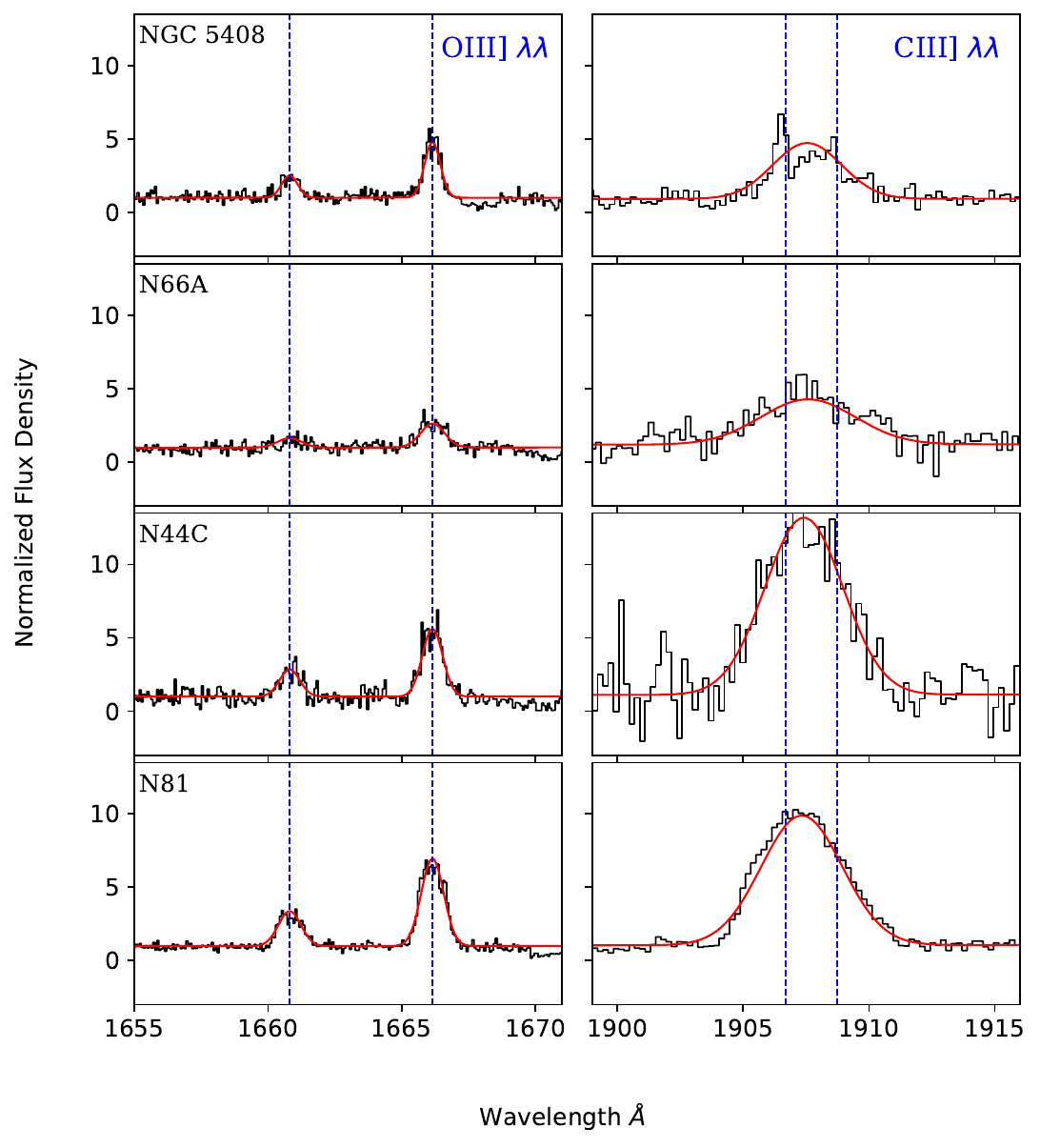"}
    \caption{Rest frame HST/COS spectra of \Ciiiuv~$\lambda\lambda$1907,1909 and \Oiiiuv~$\lambda\lambda$1660,1666 emission lines. The spectra are binned by 6 pixels and scaled arbitrarily for display purposes. 
    Dashed blue lines show the expected wavelengths of each transition, and best fit Gaussian profiles to each emission line are shown in red. NGC 5408 shows a complex \Ciiiuv\ profile, that could not be fit well with a single Gaussian.}
    \label{fig:fits}
\end{figure}

We measure the fluxes of UV \Oiiiuv\ and \Ciiiuv\ emission lines from Gaussian profile fits to the calibrated spectra. The spectra and best fits are shown in Figure~\ref{fig:fits}. \Oiiiuv\ lines were fit with a double Gaussian which used a fixed theoretical flux ratio, \Oiiiuv~$\lambda$1666/$\lambda$1661~$=2.49$ as calculated using the \textsc{PyNeb} package \citep{Luridiana2015}. Both lines are fit with the same redshift and width. The \Ciiiuv\ doublet is heavily blended due to the spatially extended nature of our targets. \Ciiiuv~$\lambda\lambda$1907,1909 lines are fit with a single Gaussian which provides an adequate fit for three sources; a double Gaussian fit gives consistent total flux and does not substantially add information given the blended spectral profile of the doublet.

The exception is NGC 5408, which exhibits a more complex \Ciiiuv~$\lambda\lambda$1907,1909 profile due to the object's substructure. Because a single Gaussian does not adequately fit for this object, we instead summed the flux over the rest-frame wavelength range 1905 to 1911 {\AA} after subtracting a best-fit continuum.

The UV observations for both of our literature sources, N88A and Mrk 71, are described in \citet{Garnett1995}. 
We used the dust-corrected  \Ciiiuv~$\lambda\lambda$1907,1909 and \Oiiiuv~$\lambda\lambda$1660,1666 emission line fluxes from \citet{Kurt99} in our analysis of N88A. For Mrk 71 (sometimes identified as NGC 2363 or NGC 2366 in the literature), we used the fluxes from \citet{Smith23}. The optical results for Mrk 71 and N88A are from \citet{Esteban2009} and \citet{Toribio2017} and are included in Table ~\ref{tab:rls}.

\subsection{Dust attenuation correction}
\label{sec:dust} 

Dust attenuation curves in the UV can vary, and the choice of curve used for attenuation correction affects the C/O abundance derived from UV emission lines. In this work, we chose an appropriate curve for each target's properties and values of C(H$\beta$) measured from \Hi\ Balmer and Paschen lines (see Table~\ref{tab:rls} for the values and sources).
For our target \Hii\ region N44C in the Large Magellanic Cloud, we adopt the average LMC2 Supershell Sample from \citet{Gordon03}.
We adopt the average SMC Bar Sample for the \Hii\ regions in the SMC (N66A and N81) and for NGC 5408. 
A dust attenuation curve has not been directly measured for NGC 5408. We use the SMC curve because NGC 5408 has a low metallicity close to that seen in the SMC. 
Additionally, we examine a secondary set of results for an LMC dust law and we find log(C$^{2+}$/O$^{2+}$) to increase by 0.1 dex, after correcting for temperature fluctuations and assuming temperatures determined from \Oiii\ (Case 2 in Section~\ref{sec:cases}). 

We used the dereddened line intensities in \citet{Kurt99} for N88A, where they determined an appropriate attenuation law for this \Hii\ region in the SMC. For Mrk 71, we used the \cite{Calzetti94} dust law since no 2175 {\AA} bump was detected \citep{Smith23}, and this result most closely matched the result using the original dust law \citep{Seaton79} applied in \citet{Garnett1995}.

The 2175 {\AA} bump strength can affect the observed flux of \Ciiiuv\ relative to \Oiiiuv, and we found that the choice of attenuation curve can significantly change the derived C$^{2+}$/O$^{2+}$. For example, for N66A, there is an increase of 0.1 dex if we instead assume an LMC or \cite{Calzetti94} dust curve, and a change of 0.17 dex under the \cite{Cardelli89} attenuation law.  For the \Hii\ regions where we applied the SMC dust curve, we found C$^{2+}$/O$^{2+}$ to be generally consistent when using the \cite{Seaton79}, Cardelli, and Calzetti laws. Under the SMC dust curve C$^{2+}$/O$^{2+}$ (Case 2) for N81 increases by 0.06 dex. Using the SMC dust curve for N44C shows a decrease of 0.1 dex.
The net effect or our sample is a systematic uncertainty of up to 0.07 dex in the mean C$^{2+}$/O$^{2+}$, considering the full range of attenuation curves.

\subsection{Ionic abundances}
\label{sec:abundances} 
To calculate the carbon and oxygen ionic abundances, we used the package \textsc{PyNeb} \citep{Luridiana2015} to calculate the emission line emissivities. The electron temperatures (\Te) and densities (n$_e$) are listed in Table~\ref{tab:rls} with their sources in the caption. 
In the case of \Te(\Oiii), we report values recalculated from the previously reported \Oiii~$\lambda$4363/$\lambda$5007 fluxes using the atomic data sets adopted in this work. We also recalculated the ADF(O$^{2+}$). We use \citet{TZ17} for O$^{2+}$ collisional strengths, \citet{FFT04} for O$^{2+}$ transition probabilities, \citet{Bal85} for C$^{2+}$ collisional strengths, and \citet{WFD96} for C$^{2+}$ transition probabilities.  

Densities for N66A, N44C, N81, Mrk 71, and N88A are determined from [\ion{Cl}{3}],  since its ionization potential ensures the chosen density traces the same ionization region as the \Oiiiuv\ and \Ciiiuv\ CELs. For NGC 5408 we used an average of the densities reported for other ions by \cite{Esteban2014}, as no density from [\ion{Cl}{3}] was available.

The derived abundances are sensitive to electron temperatures \Te, which may differ for \Oiiiuv\ and \Ciiiuv\ since they have different ionization potentials of 35 and 24 eV respectively. 
With the objective of maintaining consistency between C$^{2+}$/O$^{2+}_{CELs}$ and C$^{2+}$/O$^{2+}_{RLs}$, for our fiducial case we assume the same \Te(\Oiii) (given in Table~\ref{tab:rls}) to derive abundances from both the \Oiiiuv\ and \Ciiiuv\ emission doublets. The resulting C$^{2+}$/O$^{2+}_{CELs}$ values are listed in Table~\ref{tab:rls} and compared directly with optical RL-based measurements in Figure~\ref{fig:results}, showing excellent agreement. 
In Section~\ref{sec:cases} we assess how different assumptions for the temperature structure affect the abundances. Regardless of the chosen \Te, the uncertainty in temperature represents the dominant source of uncertainty in our measurements of C$^{2+}$/O$^{2+}$.

\section{Carbon abundance and ADF}
\label{sec:results} 

We now discuss the abundance of carbon relative to oxygen, and the extent to which it is robust under various assumptions. Specifically we focus on the C$^{2+}$/O$^{2+}$ ionic abundance which is measured directly.
By comparing RL- and CEL-based measurements, we are also able to determine whether C$^{2+}$ exhibits an abundance discrepancy factor.

\subsection{$C^{2+}$/$O^{2+}$ and the Carbon ADF}
\label{sec:CtoO} 

\begin{figure}
    \label{fig:c_o}
    \includegraphics[width=\columnwidth]{"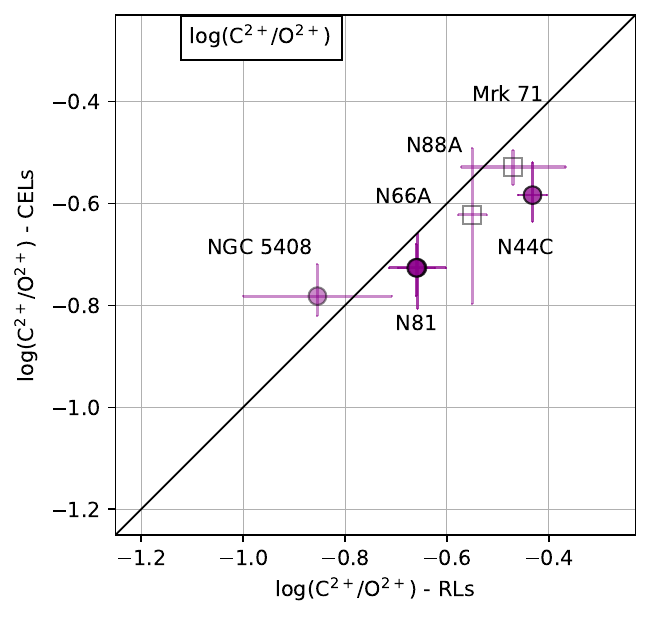"}
    \caption{log(C$^{2+}$/O$^{2+}$) derived from RLs (from \citealt{Esteban2014}, \citealt{Esteban2009}, and \citealt{Toribio2017}) versus the equivalent UV CEL-based measurements described in Section~\ref{sec:abundances}. For the UV CEL-based abundances, we used \Te(\Oiii) for both C$^{2+}$ and O$^{2+}$. 
    The UV CEL and optical RL values are in good agreement, all falling near the 1:1 line shown in black. The square symbols represent sources from the literature (Mrk 71 and N88A) which are included in our analysis. We note that N66A and N81 overlap due to their nearly identical values.
    }
    \label{fig:results}
\end{figure}

\begin{figure}
    \includegraphics[width=\columnwidth]{"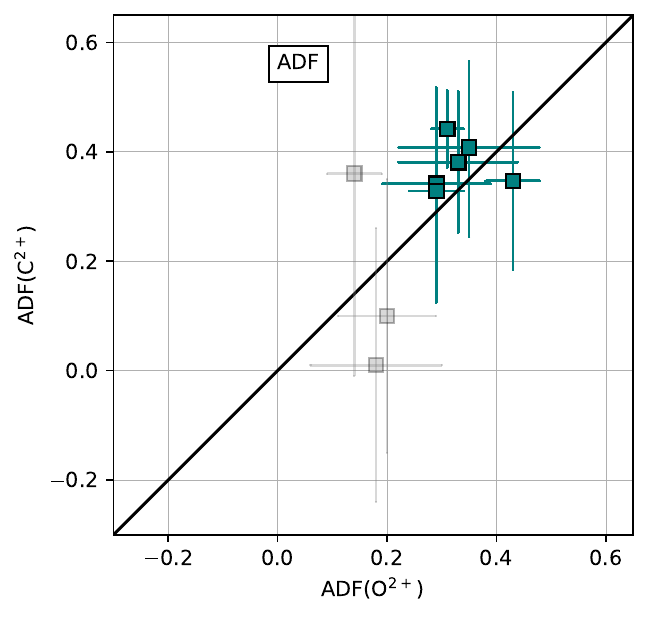"}
    \caption{The ADF(O$^{2+}$) versus ADF(C$^{2+}$) for the nearby \Hii\ regions in our sample (teal squares) and three objects from \citet{Toribio2017} (gray squares). Values for the ADF(O$^{2+}$) originate from \citet{Esteban2014} and \citet{Toribio2017}. 
    In all cases the ADF(C$^{2+}$) and ADF(O$^{2+}$) are consistent, illustrated by the 1:1 line in black.} 
    \label{fig:adf}
 
\end{figure}

The abundance ratios log(C$^{2+}$/O$^{2+}$) derived from UV CELs with HST/COS are shown in Figure~\ref{fig:results}, compared to the equivalent abundance ratio derived from optical RLs. We include literature measurements of the objects N88A and Mrk 71 \citep{Esteban2009,Toribio2017,Smith23,Kurt99} as squares. The CEL- and RL-based measurements agree well for the sample, with an average difference of $0.05\pm0.03$~dex. The individual \Hii\ regions in our sample all agree to within $<0.1$ dex.

The excellent agreement of C$^{2+}$/O$^{2+}$ shown in Figure~\ref{fig:results} is especially striking given the well-known abundance discrepancy factor for O$^{2+}$. Specifically, O$^{2+}$ abundances based on optical CELs are consistently lower than from RLs, with differences of 0.29--0.43 dex for our sample (Table~\ref{tab:rls}). Our targets show somewhat larger discrepancies ($\sim$0.3--0.4 dex) than the average oxygen ADF of $\sim$0.25 dex found in larger samples \citep[e.g.,][]{Peimbert2017}, which makes the agreement in Figure~\ref{fig:results} even more notable. The consistency of C$^{2+}$/O$^{2+}$ measurements implies that C$^{2+}$ must exhibit a similar ADF as O$^{2+}$. 

We can now directly measure the ADF of C$^{2+}$ from our UV CEL results in comparison with previous RL measurements. We define the ADF of an ion $X$, measured relative to H$^+$ abundance in an \Hii\ region, as
\begin{equation}
\label{eq:ADF}
    ADF(X)= \log(X/H^+)_{RLs}-\log(X/H^+)_{CELs}
\end{equation}
where the CEL abundance is measured from the \Te\ method assuming $t^2 = 0$ (i.e., no temperature fluctuations).
We adopt logarithmic units whereas some other studies present the linear factor (i.e., $10^{ADF}$ as defined here). The ADF(O$^{2+}$) is established for our targets via Equation~\ref{eq:ADF} based on optical spectroscopy. For the carbon ADF, we use (C$^{2+}$/O$^{2+}$)$_{CELs}$ instead of a direct (C$^{2+}$/H$^{+}$)$_{CELs}$, as the former is measured with relatively little uncertainty from our UV spectra. This allows a more precise result via
\begin{equation}
\label{eq:ADF_C}
\begin{split}
    ADF(C^{2+}) = & \log(C^{2+}/O^{2+})_{RLs} - \log(C^{2+}/O^{2+})_{CELs} \\
    & + ADF(O^{2+}).
\end{split}
\end{equation}
Each individual term in Equation~\ref{eq:ADF_C} is derived from either UV-to-UV or optical-to-optical line ratios. This approach minimizes systematic error from dust attenuation, aperture matching, absolute flux calibration, and other effects which can arise when combining data from different instruments and across large wavelength separations. From Equation~\ref{eq:ADF_C} it is clear that if the abundance discrepancies are similar (i.e., $ADF(C^{2+}) \approx ADF(O^{2+}$)), we can expect the C$^{2+}$/O$^{2+}$ from CELs and RLs to likewise be similar. This is the case for all targets in our sample, given the measurement uncertainties.

The ADF(C$^{2+}$) values for our targets calculated using Equation~\ref{eq:ADF_C} are listed in Table~\ref{tab:rls}. 
In Figure~\ref{fig:adf} we plot ADF(C$^{2+}$) from this work compared with previously reported measurements of ADF(O$^{2+}$). We find clear evidence that carbon exhibits a non-zero ADF, with an unweighted sample mean ADF(C$^{2+}$) of $0.40\pm0.02$ dex.

There is no statistically significant correlation between the ADF(C$^{2+}$) and ADF(O$^{2+}$) for the six objects in our sample or when we include the additional three objects from \citet{Toribio2017}, represented as the gray squares in Figure ~\ref{fig:adf}. This is likely due to the limited sample size and a limited dynamic range in the ADFs of our objects. The ADFs are consistent within $1\sigma$ on average for our fiducial assumptions described in Section~\ref{sec:Analysis}, and within $2\sigma$ for the sample means. These results suggest that the C/O abundance is relatively robust across the optical RL and UV CEL methods considered in this work.

\subsection{Temperature structure}
\label{sec:cases}

In this section we assess how the UV-based C$^{2+}$/O$^{2+}$ abundance changes under different assumptions for nebular temperature structure. In all cases the high-ionization zone temperature \Te(\Oiii) is used for O$^{2+}$ abundances. The energy required to further ionize O$^{2+}$ is 55 eV, higher than the 48 eV required to ionize C$^{2+}$. We determine a secondary ionization zone for C$^{2+}$ using \Te\ measured from \Siii\ emission \citep[e.g.,][]{Mingozzi2022}. Though S$^{2+}$ and C$^{2+}$ have very similar ionization potentials (23 and 24 eV), the energy to ionize them further differs (35 and 48 eV).
As a result, \Ciiiuv\ CEL emission is expected to be distributed across both S$^{2+}$ and O$^{2+}$ regions. For C$^{2+}$ we therefore considered two scenarios: a high-ionization zone (\Te(\Ciiiuv) = \Te(\Oiii)), and a case between the high and intermediate ionization zone traced by \Siii\ in which \Te(\Ciiiuv) $= \frac{1}{2}$(\Te(\Oiii) + \Te(\Siii)). We refer to the latter as \Te$_{,mid}$. \citet{Esteban2009} did not report \Te(\Siii) for Mrk 71, so it is not represented in cases using \Te$_{,mid}$ for \Ciiiuv. For both \Oiiiuv\ and \Ciiiuv\ we consider two limiting cases of temperature fluctuations, with $t^2=0$ or the $t^2$ values reported in Table~\ref{tab:rls}. 

\begin{figure*}
\centering

\includegraphics[width=17cm]{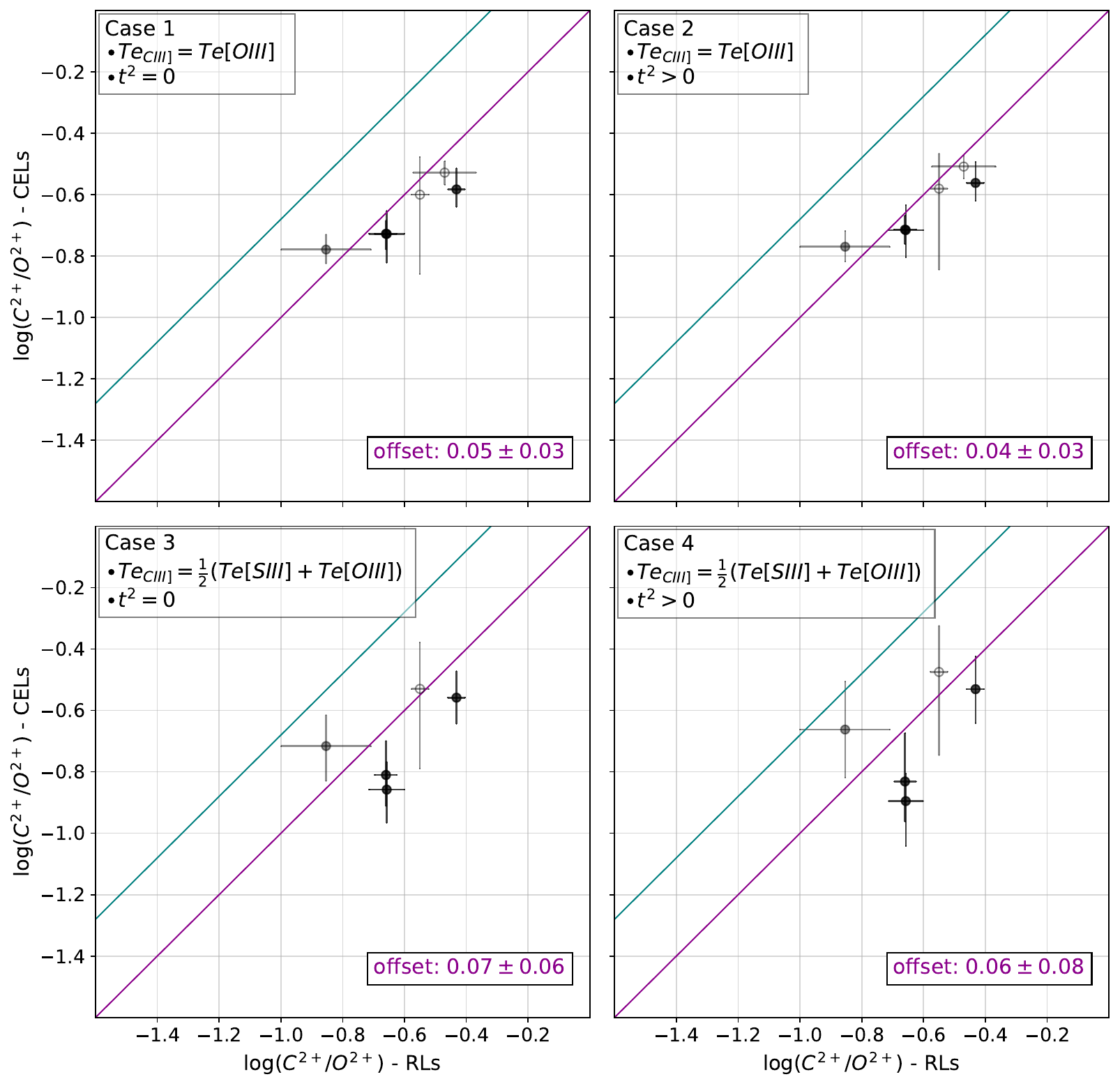}
\caption{$C^{2+}$/O$^{2+}_{CELs}$ versus $C^{2+}$/O$^{2+}_{RLs}$ for the four cases outlined in Section ~\ref{sec:cases} and summarized in each panel. t$^2$, \Te([\Oiiiuv), $C^{2+}$/O$^{2+}_{RLs}$, and \Te(\Siii) are from the iterature (see Table~\ref{tab:rls}). The black points represent the four \Hii\ regions presented and the open points are from the literature. A 1:1 relationship is indicated in purple with all cases showing good agreement. The average offset between the ionic abundances obtained from RLs and CELs is displayed in the bottom right of each panel. The teal line is shifted by the average ADF($O^{2+}$) of our \Hii\ regions, showing the trend we would expect if Carbon did not exhibit an ADF.}
\label{fig:CtoO}
\end{figure*}

In total we examine four different cases (varying \Te(\Ciiiuv) and $t^2$) described below. The  t$^2$ values used here are from literature sources (see Table~\ref{tab:rls}) and are derived from the formalism in \citet{peimbert1969}. We apply the same t$^2$ values to calculate the C$^{2+}$/O$^{2+}$ abundance. However, the t$^2$ derived from direct integration \cite[e.g.,][]{Chen2023} is slightly different and would lead to a systematic $\sim$0.04 dex increase in the $C^{2+}$/O$^{2+}$ results. This change is relatively small and does not impact our conclusions.
The results are illustrated in Figure~\ref{fig:CtoO}. 
We report the C$^{2+}$/O$^{2+}$ values for each case in Table~\ref{tab:CtoO}. The average offset from the RL-based results is also listed in Table~\ref{tab:CtoO}, which we calculate as the unweighted mean difference of CEL and RL values. Our fiducial assumptions correspond to Case 1.

\begin{table*}
 \caption{C$^{2+}$/O$^{2+}_{CELs}$ measured from UV CELs for each of the four cases described in Section~\ref{sec:cases}, for various assumptions about the temperature. Case 1 represents our fiducial assumptions of \Te(C III]) = \Te([O III]) and $t^2 = 0$. The Case 1 results are also reported in Table~\ref{tab:rls}; we include them here for ease of comparison. The right-most column is the average offset of C$^{2+}$/O$^{2+}_{CELs}$ listed in this table compared with RLs from optical spectra (Table~\ref{tab:rls}), calculated as an unweighted mean. The different assumed temperature structure has a minor effect, and in all cases the CEL- and RL-based abundances agree within $<2\sigma$.}

 \label{tab:CtoO}
 \centering
 \begin{tabular}{ccccccccc}
 \toprule
 \textsf{} & \multicolumn{1}{c}{\textsf{NGC 5408}} & \multicolumn{1}{c}{\textsf{N66A}}& \multicolumn{1}{c}{\textsf{N44C}}& \multicolumn{1}{c}{\textsf{N81}}& \multicolumn{1}{c}{\textsf{N88A}}& \multicolumn{1}{c}{\textsf{Mrk 71}}\\
  & $C^{2+}/O^{2+}$  & $C^{2+}/O^{2+}$  & $C^{2+}/O^{2+}$  & $C^{2+}/O^{2+}$ & $C^{2+}/O^{2+}$ & $C^{2+}/O^{2+}$ & average offset\\ 
 \hline
Case 1 & -0.78$\pm0.05$ & -0.73$^{+0.07}_{-0.10}$ & -0.58$\pm0.06$ & -0.73$\pm0.05$ & -0.61$^{+0.12}_{-0.18}$ & -0.52$\pm0.04$ & 0.05$\pm$0.03 \\
Case 2 & -0.77$\pm0.05$ & -0.72$^{+0.07}_{-0.09}$ & -0.56$\pm0.06$ & -0.71$\pm0.06$ & -0.59$^{+0.12}_{-0.19}$ & -0.51$\pm0.04$ & 0.04$\pm$0.03 \\
Case 3 & -0.72$\pm0.10$ & -0.87$\pm0.09$ & -0.56$\pm0.09$ & -0.80$^{+0.11}_{-0.13}$ & -0.54$^{+0.12}_{-0.19}$ & -0.46$\pm0.07$ & 0.07$\pm$0.06 \\
Case 4 & -0.67$^{+0.16}_{-0.14}$ & -0.91$\pm0.11$ & -0.54$\pm0.11$ & -0.82$\pm0.15$ & -0.48$^{+0.12}_{-0.21}$ & -0.41$\pm0.10$ & 0.05$\pm$0.08 \\

\end{tabular}
\end{table*}

\begin{enumerate}
  \item \Te(C III]) = \Te([O III]), $t^2 = 0$:\\
  Case 1 employs \Te(\Oiii) to determine both carbon and oxygen ionic abundances without correcting for temperature fluctuations ($t^2 = 0$). This is our fiducial case corresponding to the results described in previous sections.
  As shown in Figure~\ref{fig:CtoO}, this method yields an average offset of $0.05\pm0.03$ from the UV CELs relative to $C^{2+}$/O$^{2+}_{RLs}$. The agreement between this method and the results obtained from recombination lines supports the validity of using \Te(\Oiii) without correcting for temperature fluctuations when determining C/O.

  \item \Te(C III]) = \Te([O III]), $t^2 > 0$:\\
  Case 2 uses \Te(\Oiii) for both carbon and oxygen ionic abundances, with temperature fluctuations corrections applied to both. This approach gives the closest agreement to results found with recombination lines. 
  Consequently, we conclude that this is the best method to obtain C$^{2+}$/O$^{2+}$ values from UV CELs which are consistent with RLs.
  However, it is also consistent (within $\sim$0.01 dex) with Case 1, demonstrating that temperature fluctuations have relatively little effect on the derived C/O abundance.

  \item \Te(C III]) = \Te$_{,mid}$, $t^2 = 0$:\\
  The average of \Te(\Siii) and \Te(\Oiii) (denoted as T$_{e,mid}$) was used for the \Ciiiuv\ temperature in Case 3, while \Te(\Oiii) was used for oxygen. Similar to Case 1, no correction for temperature fluctuations is applied. The results are broadly consistent with those of Case 1 in which \Te(\Oiii) is used for \Ciiiuv, and are also consistent with C$^{2+}$/O$^{2+}$ from optical RLs.
  However, the scatter is somewhat larger compared to Case 1.

  \item \Te(C III]) = \Te$_{,mid}$, $t^2 > 0$:\\
  Case 4 uses T$_{e,mid}$ for \Ciiiuv, with temperature fluctuation corrections applied to both carbon and oxygen. The C$^{2+}$/O$^{2+}$ ratio generally agrees with with the results found for RLs and other cases, though Case 4 has the largest scatter.

\end{enumerate}

An ionization correction factor (ICF) is applied to C$^{2+}$/O$^{2+}$ to obtain the total abundance ratio of C/O. Thus, any bias in C$^{2+}$/O$^{2+}$ directly translates into the same amount of bias in C/O. Analysis of these four cases shows that the UV-based C$^{2+}$/O$^{2+}$ (and hence total C/O) abundance is robust, as long as the same temperature fluctuations are adopted for both ions.  Applying the same $t^2$ to both \Ciiiuv\ and \Oiiiuv\ gives similar abundances to those obtained assuming $t^2 = 0$, with an average difference of 0.01 dex. Adopting T$_{e,mid}$ cf. \Te(\Oiii) for the \Ciiiuv-based abundance produces a similarly modest change of 0.02 dex on average. In all four cases above, the C$^{2+}$/O$^{2+}$ derived from UV CELs is within 0.04--0.07 dex of the RL-based values on average, and statistically consistent ($\lesssim 1\sigma$).

\subsection{Implications for UV-based nebular abundances and the ADF}

The results of this work allow us to assess the reliability of UV CELs to measure nebular abundance patterns. Applications include star forming galaxies at high redshifts ($z\simeq2$--10; e.g., \citealt{Christensen2012,Cordova2022,Jones23,D'Eugenio2024,Hayes2024, Hu24}), and low-metallicity dwarf starbursts in the local universe \citep[e.g.,][]{Berg2019}. The data quality and available measurements of physical properties vary widely. For example, \Te(\Siii) and ADF measurements (and derived $t^2$) are often not practical, especially at high redshifts. Relatedly, abundance pattern measurements in the literature adopt various different assumptions about nebular temperature structure. 
The cases discussed in Section~\ref{sec:cases} offer guidance on interpreting such measurements based on different underlying assumptions.

From a pragmatic standpoint,  our four cases are mutually consistent ($<1\sigma$) and within $\leq0.03$ dex in C$^{2+}$/O$^{2+}$. The main difference is that assuming a value $t^2 > 0$, or otherwise correcting for the ADF, results in higher total abundances of C$^{2+}$/H$^{+}$ and O$^{2+}$/H$^{+}$. For purposes of C/O abundance, Case 1 is the simplest application. Our results support using this approach, namely assuming \Te(\Ciiiuv) = \Te(\Oiii) and $t^2 = 0$ to calculate C/O abundances. However, it is clear from our analysis that an ADF correction should be adopted when comparing C/H and O/H abundances from UV CELs with those derived from RLs or other methods. Case 2 gives the best agreement in terms of combined C/O, C/H, and O/H, with the smallest offset and sample scatter out of the four cases.

Overall we conclude that C/H abundances should be measured with a similar method as for O/H. Our analysis indicates that values reported in the literature with various assumptions (such as the commonly used Cases 1 and 3, or similar) can be compared reliably, with systematic offsets $\lesssim$0.07 dex.
Likewise our analysis supports that ADF(C$^{2+}$) $\simeq$ ADF(O$^{2+}$) (within $-0.05 \pm 0.03$ dex, cf. $\sim$0.3--0.4 dex for the ADF in our sample). This is effectively often assumed and we have now quantified it with good precision.

Notably, the consistent C/O abundances measured from RLs and CELs provide insight into the possible physical origins of the ADF. If temperature fluctuations are responsible for the ADF, C$^{2+}$ must be subject to fluctuations of a similar magnitude and have a comparable $t^2$ to O$^{2+}$.
If chemical inhomogeneities are the cause, we can place constraints on the origin of the metal-rich, hydrogen-poor clumps. Since both ADFs are consistent, the C/O ratio in those cooler metal-rich regions must be similar to that in the hotter regions traced by CELs (i.e., C$^{2+}$/O$^{2+} \simeq 0.2$--0.3 for our sample). This rules out ejecta from low- and intermediate-mass stars as a main source of metallicity variations that could cause the ADF, since their C/O values are higher. The agreement between C/O derived with RLs and CELs can also be used to assess potential effects of fluorescence. $C^{2+}$/O$^{2+}_{RLs}$ is determined using \ion{C}{2} 4267 {\AA} and the O II 4650 Å multiplet. It has been shown that \ion{C}{2} 4267 {\AA} has no significant contribution from fluorescence \citep{Reyes2024}. The consistency between $C^{2+}$/O$^{2+}_{RLs}$ and $C^{2+}$/O$^{2+}_{CELs}$ suggests that fluorescent contribution to the \ion{O}{2} 4650 {\AA} multiplet is also negligible, in agreement with previous studies \citep[e.g.,][]{Storey2017, Escalante2012}. Thus, fluorescence does not contribute significantly to the ADF.

Overall our results represent a significant step for chemical abundance studies by establishing the nebular ADF\footnote{While we have discussed the ADF correction in the context of nebular abundances, such a correction should presumably also be applied for comparisons with stellar and interstellar absorption abundances, along with differences in dust depletion \citep[e.g.,][]{Jenkins2017,Ritchey2023}.} of carbon and the ability to accurately compare nebular C/O abundance measurements across cosmic time.

\section{Conclusions}
\label{sec:conclusions}

We report UV spectroscopy of four nearby \Hii\ regions with HST/COS. We measure C$^{2+}$/O$^{2+}$ abundance ratios from the \Oiiiuv\ and \Ciiiuv\ emission lines, and compare with the same ratio derived from optical recombination lines. Combined with two additional \Hii\ regions with suitable data in the literature \citep{Esteban2009,Toribio2017,Smith23,Kurt99}, we analyze a total sample of six objects, tripling the available sample with reliable UV-based measurements. 
We examine whether C$^{2+}$ exhibits an ADF similar to the well-known discrepancy for O$^{2+}$. Consequently we consider the reliability of C/H and C/O abundances obtained from different methods (i.e., RLs and CELs). 
Our main findings are as follows.

\begin{itemize}
\item Measurements of C$^{2+}$/O$^{2+}_{CELs}$, using \Te(\Oiii) to derive abundances from both the \Oiiiuv\ and \Ciiiuv\ emission lines, show good agreement with C$^{2+}$/O$^{2+}_{RLs}$.
Correcting for temperature fluctuations based on the O$^{2+}$ ADF slightly improves the agreement between the two methods ($0.04\pm0.03$ cf. $0.05\pm0.03$ dex).
\item The relative temperatures of \Oiiiuv\ and \Ciiiuv\ represent a potential uncertainty in the UV-based abundance method, which may increase the offset between the C$^{2+}$/O$^{2+}$ ratio derived from CELs and RLs. 
However, the results remain broadly consistent when we adopt an intermediate ionization zone temperature for \Ciiiuv.
\item We find a clear abundance discrepancy factor (ADF) for C$^{2+}$. The average C$^{2+}$ ADF of 0.40$\pm$0.02 is comparable to the average O$^{2+}$ ADF of 0.35$\pm$0.02 for our \Hii\ region sample. The ADFs are consistent within $\lesssim$~2$\sigma$ (with a difference corresponding to the offset in C$^{2+}$/O$^{2+}$ from the UV CEL and optical RL measurements, of $0.05\pm0.03$ dex for our fiducial case). This suggests that the same physical processes which cause the well-known ADF for O$^{2+}$ are likely also affecting the carbon emission lines. Additionally, an ADF correction should be adopted for C/H abundances in the same way as for O/H abundances.
\end{itemize}

\vspace{1ex}
These results provide a foundation for direct comparison of C/O abundances measured from different methods, namely optical RLs and UV CELs. This is particularly valuable to span a wide range in metallicity and cosmic time: RLs are most commonly available at high metallicity and limited to $z\sim0$, while UV CEL samples probe lower metallicity and a wide range of redshifts. Encouragingly, we find that straightforward applications of both methods (e.g., regardless of whether temperature fluctuation corrections are applied) give consistent results for C/O within our sample of six \Hii\ regions. The average difference of $0.05\pm0.03$~dex represents the possible systematic uncertainty between the two methods, assuming the standard \Te(\Oiii)-based approach.

Our results are especially important for using C/O abundances as a ``chemical clock'' to study galaxy formation at high redshifts \citep[recently reaching $z>8$; e.g.,][]{Cordova2022, Jones23, Hu24}, although we reiterate that absolute abundances such as O/H and C/H must still be treated with caution to account for the ADF.
The combination of JWST/NIRSpec and powerful ground-based observatories have the capability to measure C/O from the rest-frame UV \Ciiiuv~$\lambda\lambda$1907,1909 and \Oiiiuv~$\lambda\lambda$1660,1666
lines spanning redshifts $z\simeq 2$--10. We have confirmed that these measurements can be directly compared on the same scale with those based on optical RLs at $z\simeq0$, and quantified the uncertainty as well as the effects of nebular temperature structure.

Looking forward, increasing the available sample size and dynamic range of C/O, temperature \Te, and ADF values will help to more precisely determine the offset between the C$^{2+}$/O$^{2+}_{CELs}$ and C$^{2+}$/O$^{2+}_{RLs}$ abundance measurements. This can be achieved both by securing UV spectroscopy of a larger sample following the methods in this paper, and with deep optical spectroscopic followup of targets with suitable archival UV data \citep[e.g.,][]{Berg2022} to measure C$^{2+}$/O$^{2+}_{RLs}$ and ADF(O$^{2+}$).\\

\section*{ACKNOWLEDGMENTS}
Based on observations with the NASA/ESA Hubble Space Telescope obtained from the Mikulki Archive for Space Telescopes at the Space Telescope Science Institute, which is operated by the Association of Universities for Research in Astronomy, Incorporated, under NASA contract NAS5-26555. Support for Program number HST-GO-16697 was provided through a grant from the STScI under NASA contract NAS5-26555. 
TJ acknowledges support from the NASA under grant 80NSSC23K1132, and from a UC Davis Chancellor's Fellowship. PK acknowledges support from the UC Davis Dean's Distinguished Graduate Fellowship. YC is supported by the Direct Grant for Research (C0010-4053720) from the Faculty of Science, the Chinese University of Hong Kong.

%

\vspace{5mm}
\facilities{HST(COS)}





\bibliographystyle{aasjournal}
\bibliography{sample631}



\end{document}